\documentclass[review,5p,times,authoryear]{elsarticle}

\usepackage{amssymb}
\usepackage{amsmath}
\usepackage{booktabs}
\usepackage{multirow}
\usepackage[colorlinks=true, citecolor=blue]{hyperref}
\usepackage{float}

\begin{document}

\begin{frontmatter}

\title{Robust Subgroup Method Using DE Algorithm for Resonance Self-Shielding Calculation} 

\author[1]{Beichen Zheng}
\author[2]{Ying Chen}
\author[3]{Lili Wen\corref{cor1}}
\author[4]{Xiaofei Wu}

\affiliation{organization={China Nuclear Data Center, China Institute of Atomic Energy},
	city={Beijing},
	country={China}}

\begin{abstract} 
	This paper presents an enhanced version of the subgroup method for resonance self-shielding treatment, termed the robust subgroup method, which integrates Robust Estimation (RE) with a Differential Evolution (DE) algorithm. The RE approach is employed to handle model misspecification and data contamination, while the DE algorithm serves as an optimization tool within the RE framework to obtain constrained solutions. Numerical validation against experimental benchmarks shows that the proposed method removes a systematic absorption bias in conventional subgroup fits that would otherwise depress reactivity. This bias appears only in benchmarks sensitive to \({}^{238}\mathrm{U}\). Mechanistically, it reflects a threshold-like conditioning failure: strong self-shielding leverage dominates the loss and is magnified by dilution-induced multicollinearity. This adverse conditioning appears to be seeded by a narrow, sparse resonance structure at low energies in fertile even-even nuclides, thereby causing rapid self-shielding response saturation and a weak Doppler broadening. By bounding influence and enforcing feasibility within an RE-DE framework, the inferred subgroup parameters track the underlying physics more faithfully, improving the predictive fidelity of subsequent transport simulations.
\end{abstract}



\end{frontmatter}



\section{Introduction}
Resonance self-shielding is essential in lattice physics because resonant cross-section structure depresses the local flux and thus strongly affects transport and burnup results. The objective of resonance self-shielding calculation is to produce effective multigroup cross sections in resonance regions. The most direct and accurate method is the ultra-fine group (UFG) method \cite{ishiguro1971peaco,sugimura2007resonance}, which discretizes the resonance energy range into an extremely fine grid and solves the slowing-down equation to preserve individual resonance shapes and interference. However, the method's high computational and memory demands, together with complex preprocessing, make it unsuitable for many repeated or large-scale problems. Another traditional method is the equivelance theory \cite{askew1966general,stamm1983methods},  which is computationally efficient and fits naturally into multigroup workflows, but it relys on homogenization and often fails to capture spatially dependent self-shielding, streaming and complex interference without supplemental corrections like the Dancoff factor. The subgroup method \cite{levitt1972probability,cullen1974application,nikolaev1976comments} has strong geometric flexibility and offers a practical balance between efficiency and accuracy. By representing each coarse group with a few cross-section levels and weights and solving subgroup fixed-source problems,  it captures much of the spatial, angular and interference behavior at a fraction of the UFG cost. Its ability to adapt weights and levels to local moderator-fuel arrangements and streaming paths makes it particularly effective in heterogeneous geometries, but achieving UFG-level accuracy still requires careful generation of subgroup parameters and explicit treatment of resonance interference. 

The subgroup method, also known as the probability table method, encompasses two main approaches for determining the subgroup parameters \cite{hebert2002computing}. The first approach is the mathematical probability table method \cite{cullen1974application,ribon1986probability}, which is based on Gauss quadrature and utilizes the cross-section moment conservation principle in the direct processing of pointwise cross-section data. The non-integer order cross-section moments ranging from -1 to 0 are chosen to be conserved, since they are found to be related to the effective cross-sections where the background cross sections span from 0 to infinity \cite{chiba2006improvement}. The second approach, the physical probability table method \cite{casal1991helios}, preserves the effective multi-group cross sections derived from exact slowing down calculations with chosen diluents. Despite the subgroup parameters varying depending on the reactor type, this method requires fewer energy groups. Within the framework of the physical probability table method, the constrained over-determined problem formulated via the least-squares can be effectively solved using metaheuristic algorithms \cite{safarzadeh2015resonance,li2021improved}. However, while ordinary least squares (OLS) offers a convenient baseline for the overdetermined fits used to reconstruct effective cross sections, its quadratic loss makes it brittle in the regimes of interest here. A small number of energy-temperature-background combinations can yield disproportionately large residuals due to (i) violations of modeling assumptions (e.g., surrogates of slowing-down physics, ill-conditioning across background grids,  leverage-prone residual structure) and (ii) data contamination in a broad sense (heavy-tailed errors, numerical artifacts in reference calculations, and sporadic outliers). In such cases, OLS tends to overreact to extreme deviations and propagate them into subgroup levels and weights. Robust estimation \cite{huber1964robust,hampel1968contributions} is designed precisely to mitigate these failure modes by replacing the squared loss with a bounded-influence objective, limiting the impact of assumption violations and anomalous data while preserving near-OLS behavior where the model is well specified. In what follows, a scale-preestimated, redescending M-estimator is adopted and embedded in a Differential Evolution (DE) \cite{storn1995differential,osaba2021tutorial} solver to handle the resulting constrained, nonlinear optimization efficiently. 
This paper first analyzes the subgroup method, identifying the root causes of systematic discrepancies and limitations inherent in the conventional OLS approach. To overcome these shortcomings, we introduce a RE framework centered on a  M-estimator. The practical implementation of this framework is achieved using a DE algorithm tailored for constrained optimization. Finally, we present and discuss numerical results, validating them against experimental benchmarks to assess the performance and accuracy of the proposed method.

\section{Subgroup method} \label{Subgroup method}

Accurately modeling resonance self-shielding---the significant flux depression caused by resonances---is critical in lattice physics. The subgroup method provides an efficient and accurate solution.
The method's integration strategy partitions the complex resonance structure by cross-section magnitude, rather than energy. This is physically motivated because flux depression correlates strongly with the cross-section value itself. This approach transforms the complex, oscillatory cross-section behavior within a coarse energy group into a discrete "probability table". This table consists of a few subgroups, each characterized by a representative subgroup level ($\sigma_n$, representing a cross-section value) and a subgroup weight ($\omega_n$, representing the probability or fraction of the energy group associated with that level). The effective multigroup cross-section ($\sigma_{xg}$) for reaction type $x$ in group $g$ is then calculated as a flux-weighted average over these $N$ subgroups:
\begin{equation}
\begin{aligned}
  \sigma_{x,g} &=
  \frac{\int_{u_{g+1}}^{u_g} \sigma_x(u)\,\phi(u)\,du}{\int_{u_{g+1}}^{u_g} \phi(u)\,du} \\
  &= \frac{\int_{\sigma_{\min,g}}^{\sigma_{\max,g}} \sigma_x \,\phi(\sigma)\,\omega(\sigma)\,d\sigma}
       {\int_{\sigma_{\min,g}}^{\sigma_{\max,g}} \phi(\sigma)\,\omega(\sigma)\,d\sigma} \\
  &= \frac{\sum_{n=1}^{N} \sigma_n \phi_n \omega_n}{\sum_{n=1}^{N} \phi_n \omega_n}
\label{ref2_latex}
\end{aligned}
\end{equation}
where $u = \ln\left({E_0}/{E}\right)
\label{eq:lethargy_def}$ is the lethargy, $\omega_n$ is the subgroup weight for the $n$-th subgroup, $\sigma_n$ is the subgroup level for the $n$-th subgroup, and $N$ is the number of subgroups. Together, the subgroup levels and weights are referred to as the subgroup parameters.

To calculate the effective cross-sections for each resonance group, the scalar flux for each subgroup is required. It can be obtained by solving the fixed-source transport equation for subgroup level \(n\), as shown in the following equation:
\begin{equation}
    \Omega \cdot \nabla \Phi_n + \left( N_r \sigma_{a,n} + \sum_i \lambda_i N_i \sigma_{i,p} \right) \Phi_n = \sum_i \lambda_i N_i \sigma_{i,p}
\end{equation}
where \( \Phi_n \) is the scalar flux for the \(n\)-th subgroup level,
\( N_r \) is the atomic number density of the resonant isotope,
\( \sigma_{a,n} \) is the microscopic absorption cross-section for the resonant isotope at the \(n\)-th level,
\( \lambda_i \) is the intermediate resonance parameter at energy group \(g\) for nuclide \(i\),
\( N_i \) is the atomic number density for nuclide \(i\),
\( \sigma_{i,p} \) is the microscopic potential cross-section for nuclide \(i\),
\( \Omega \) is the direction vector of the neutron flux. This equation is solved using a transport solver to compute the flux distribution for each subgroup level. Once the fluxes are determined, the effective resonance cross-sections can be obtained for each broad group and resonance nuclide. 

To account for inter-isotopic resonance interference, the Bondarenko iteration method \cite{Stammler1998,yin2025study,zhao2025research} is commonly applied. This iterative procedure updates the effective cross sections by accounting for the background contribution from other resonant isotopes. For a given nuclide $i$, the absorption background contributed by all other isotopes $j$ can be expressed as
\begin{equation}
	\sigma_{x,i} = \frac{1}{N_i} \sum_{j \ne i} N_j ,\sigma_{a,j}
	\label{eq:sx}
\end{equation}
where $N_j$ is the number density of isotope $j$ and $\sigma_{a,j}$ is its effective absorption cross section.
The effective absorption cross section for nuclide $i$ is then approximated by
\begin{equation}
	\sigma_{a,i} =
	\frac{\displaystyle \sum_{n} w_{n,i}\frac{\sigma_{n, i}\sigma_{b, n, i}}{\sigma_{n i} + \sigma_{x, i} + \sigma_{b, n, i}}}
	{\displaystyle \sum_{n} w_{n, i}\frac{\sigma_{b, n, i}}{\sigma_{n, i} + \sigma_{x, i} + \sigma_{b, n, i}}}
	\label{eq:sai}
\end{equation}
where $n$ indexes the subgroups, $w_{n, i}$ are the subgroup weights for isotope $i$, $\sigma_{n, i}$ are the subgroup microscopic absorption cross sections, and $\sigma_{b, n, i}$ denotes the subgroup background cross section for isotope $i$.
Equations~\eqref{eq:sx} and \eqref{eq:sai} show that $\sigma_{a, i}$ and $\sigma_{a, j}$ are mutually dependent through $\sigma_{x, i}$. This coupling is resolved by a fixed-point iteration: starting from an initial guess (e.g., infinite-dilution cross sections), iteratively update the self-shielded flux and the effective cross sections using Eqs.~\eqref{eq:sx} and \eqref{eq:sai} until the results converge. This yields a self-consistent treatment of resonance self-shielding with inter-isotopic interference.

\subsection{ The determination of subgroup parameters }

Calculating resonance region flux $\phi(E)$ is challenging due to self-shielding from sharp peaks. Two limiting approximations simplify this: The Narrow Resonance (NR) approximation assumes resonances are narrow compared to moderator collision energy loss, meaning neutrons effectively bypass them, preserving the asymptotic $1/E$ flux shape. Conversely, the Wide Resonance (WR) approximation assumes resonances are broad compared to resonant nuclide collision energy loss, making the flux follow the resonance profile, often as $\phi(E) \propto 1/\Sigma_t(E)$.
However, many significant resonances, particularly important low-energy ones in heavy nuclides like $^{238}$U, fall between these extremes, making neither the pure NR nor WR approximations fully accurate. The Intermediate Resonance (IR) approximation \cite{goldstein1962theory} was developed specifically to address this intermediate regime. It bridges the gap by interpolating between the NR and WR limits using the Goldstein-Cohen factor, $\lambda$. This factor, typically ranging between 0 (representing the WR limit) and 1 (representing the NR limit), quantifies the nature of scattering collisions within the resonance and generally depends on the nuclide, energy group, and temperature.

The IR approximation models the scattering source as a $\lambda$-weighted sum of NR and WR  components. Solving the slowing-down equation with this source yields the standard IR flux expression at lethargy $u$:
\begin{equation}
\begin{split}
\phi(u) &= \frac{\sum_i \lambda_i \Sigma_{p,i}}{\Sigma_a(u) + \sum_i \lambda_i (\Sigma_s(u) - \Sigma_{p,i}) + \sum_i \lambda_i \Sigma_{p,i}} \\
&= \frac{\sigma_b}{\sigma_a(u) + \lambda_r \sigma_{rs,r}(u) + \sigma_b}
\end{split}
\label{eq:IR_flux_full_concise_revised_nou}
\end{equation}
where $\Sigma_a, \Sigma_s, \Sigma_p$ are macroscopic cross sections, $\sigma_{rs,r} = \sigma_{s,r} - \sigma_{p,r}$ is the microscopic resonance scattering cross section of the resonant nuclide $r$, $\lambda_r$ is its Goldstein-Cohen factor, and $\sigma_b = (\sum_i \lambda_i \Sigma_{p,i}) / N_r$ is the effective background cross section per resonant atom $N_r$.
For practical subgroup method implementations, the explicit resonance scattering term $\lambda_r \sigma_{rs,r}$ in the denominator of Eq.~\eqref{eq:IR_flux_full_concise_revised_nou} is frequently neglected or implicitly handled through the fitting procedures used to generate the subgroup parameters ($\sigma_n, \omega_n$) \cite{joo2009subgroup}. This leads to the commonly used flux form:
\begin{equation}
\phi = \frac{\sigma_{b}}{\sigma_{a} + \sigma_{b}}
\label{eq:IR_flux_simplified}
\end{equation}
Here, $\sigma_a$ represents the relevant microscopic absorption cross section of the resonant nuclide $r$. The parameter $\sigma_b$ is the effective background cross section per resonant atom, which determines the magnitude of self-shielding. It quantifies the total effective scattering environment experienced by the resonant nuclide, incorporating contributions from all nuclides in the mixture. Specifically, $\sigma_b$ combines the effective potential scattering contribution from the resonant nuclide itself, weighted by its Goldstein-Cohen factor $\lambda_r$, with the contribution from all non-resonant nuclides (indexed $k \neq r$), denoted as $\sigma_0$:
\begin{equation}
\sigma_{b} = \lambda_r \, \sigma_{p,r} + \sigma_{0}
\label{eq:sigmab_def_detail}
\end{equation}
where $\sigma_{p,r}$ is the microscopic potential scattering cross section of the resonant nuclide $r$. The term $\sigma_0$ represents the external dilution provided by the surrounding material matrix per resonant atom and is defined as:
\begin{equation}
\sigma_0 = \frac{\sum_{k \neq r} \lambda_k N_k \, \sigma_{p,k}}{N_r}
\label{eq:sigma0_def_detail}
\end{equation}
where $N_k$ is the number density and $\sigma_{p,k}$ is the potential scattering cross section of the non-resonant nuclide $k$. Often, for these non-resonant nuclides, it is assumed that $\lambda_k \approx 1$. 

With the neutron flux within the resonance approximated by Eq.~\eqref{eq:IR_flux_simplified}, and the background cross section defined through Eqs.~\eqref{eq:sigma0_def_detail} and \eqref{eq:sigmab_def_detail}, the effective cross section integral in Eq.~\eqref{ref2_latex} can be explicitly formulated in terms of the subgroup parameters:
\begin{equation}
	\sigma_{x,g} = \frac{\sum_{n} \sigma_{x,n} \, \omega_{n} \, \frac{\sigma_{b}}{\sigma_{a,n} + \sigma_{b}}}{\sum_{n} \omega_{n} \, \frac{\sigma_{b}}{\sigma_{a,n} + \sigma_{b}}}
    \label{ref3:sigma}
\end{equation}
The numerator in Eq.~\eqref{ref3:sigma} represents the reconstructed resonance integral for reaction type $x$ in group $g$:
\begin{equation}
R_{x,g} = \sum_{n=1}^{N} \sigma_{x,n} \, \omega_{n} \, \frac{\sigma_{b}}{\sigma_{a,n} + \sigma_{b}}
\label{eq:Rxg_explicit_concise}
\end{equation}
It approximates the reaction rate within the group, accounting for self-shielding.

The reference values, serving as the benchmark for the subsequent subgroup parameter optimization, are generated using the NJOY code by processing data from the ENDF/B-VII.0 library. The generation process begins with the reconstruction of resonance profiles from their underlying parameters, after which several key physical effects are incorporated, most notably Doppler broadening and the thermal neutron scattering effect. The crucial resonance self-shielding effect is then addressed during the group averaging of the pointwise data. This effect can be modeled by employing a weighting flux derived from the Bondarenko formalism, parameterized by the background cross section, $\sigma_{0}$. For cases requiring higher fidelity, particularly for systems with significant broad or intermediate resonances, this flux is obtained by directly solving the slowing-down equation. Systematic computation of the resonance integral and the multigroup cross sections across a wide range of temperatures and background cross sections ultimately yields a complete reference data table.
The subgroup weights and levels are then calculated by minimizing the following function, which can be written in a unified form as in \cite{kim2016subgr}:
\begin{equation}
	f(\{\omega_{x, n,t}\}, \{\sigma_{a, n}\}) = \sum_{t=1}^T \sum_{k=1}^K \frac{1}{R_{x,k,t}^2} \left( R_{x,k,t} - \sum_{n=1}^N \omega_{x,n,t}\,\sigma_{x,n}\,\frac{\sigma_{b,k}}{\sigma_{a, n} + \sigma_{b,k}} \right)^2
\end{equation}
where K is the number of background cross sections, T is the number of temperatures. Subgroup levels are treated as temperature invariant so that the same level set can be reused across temperatures, lowering the cost of the subgroup fixed-source calculations. 

\subsection{Modeling framework and limitations} \label{Modeling framework and limitations}
The subgroup method formulates $\sigma_{xg}$ as a convex combination of the subgroup levels $\{\sigma_{an}\}$. This mathematical structure imposes a constraint on the solution, requiring that the range of $\{\sigma_{an}\}$ must encompass the reference value $\sigma_{xg}$.

\textbf{Proposition.}
Let $\{\sigma_{a,n}\}\subset (0,\infty)$ and weights $\{\omega_n\}$ satisfy $\omega_n\ge 0$ and $\sum_n \omega_n=1$.
Fix $\sigma_b>0$ and define $f(a)=\frac{\sigma_b}{a+\sigma_b}$.
Consider
\begin{equation}
	\sigma_{x,g}
	=\frac{\sum_{n}\omega_n\,\sigma_{a,n}\,f(\sigma_{a,n})}{\sum_{n}\omega_n\,f(\sigma_{a,n})}
\end{equation}
Then
\begin{equation}
	\min_{n} \sigma_{a,n} \;\le\; \sigma_{x,g} \;\le\; \max_{n} \sigma_{a,n}
\end{equation}

\textbf{Proof.}
Set $\alpha_n = \omega_n f(a_n)\ge 0$ and $S=\sum_{n\in g}\alpha_n$. Since $\sigma_b>0$ and at least one $\omega_n>0$, we have $S>0$. Then
\begin{equation}
	\sigma_{xg}
	= \frac{\sum_{n\in g}\alpha_n a_n}{\sum_{n\in g}\alpha_n}
	= \sum_{n\in g} v_n a_n
\end{equation}
where 
\begin{equation}
	v_n = \frac{\alpha_n}{S} \ge 0,\ \sum_{n\in g} v_n = 1
\end{equation}
Let $m=\min_{n\in g} a_n$ and $M=\max_{n\in g} a_n$. For each $n$,
$m\le a_n\le M$. Multiplying by $\alpha_n\ge 0$ and summing over $n$ gives
\begin{equation}
	mS \;\le\; \sum_{n\in g}\alpha_n a_n \;\le\; M S
\end{equation}
Dividing by $S>0$ yields
\begin{equation}
	m \;\le\; \frac{\sum_{n\in g}\alpha_n a_n}{\sum_{n\in g}\alpha_n} \;\le\; M
\end{equation}
i.e.,
\begin{equation}
	\min_{n\in g} a_n \;\le\; \sigma_{x,g} \;\le\; \max_{n\in g} a_n
\end{equation}
Equality on the left (resp.\ right) holds if all indices with $\alpha_n>0$ have $\sigma_{a,n}= \min_n \sigma_{a,n}$ (resp.\ $\max_n \sigma_{a,n}$).
This has a direct consequence for each independent optimization problem we solve. If a universal search range is set for all energy groups in the solver, then for this approach to be viable, the chosen range must necessarily span the entire global range of the reference data. To improve the efficiency, an ideal strategy would tailor the search domain for each energy group, but determining these specific bounds a priori is non-trivial, as our analysis only provides a loose constraint on the solution's range.
This insight explains why it is often a pragmatic and robust strategy in practical calculations to employ a crude but sufficiently wide universal search domain, for instance, by setting a practical upper bound as high as $10^{10}$ barns, as suggested in \cite{li2021improved}. Nevertheless, we find this value to be conservative in our settings; the numerical tests on the benchmark cases presented in Section~\ref{Results} indicate that upper bounds of $10^5$, $10^8$, and $10^{10}$ barns yield identical transport results, suggesting that a much lower bound is already sufficient.

To study the fitting, let us consider the RI response
\begin{equation}
	f_{\mathrm{RI}}(\sigma_b; T)
	= \sum_{n=1}^{K} \omega_n(T)\,\psi_n(\sigma_b)
	\label{ref3a}
\end{equation}
\begin{equation}
	\psi_n(\sigma_b) 
	\equiv \frac{\sigma_{a,n}\,\sigma_b}{\sigma_{a,n}+\sigma_b}
	\label{ref3b}
\end{equation}
It is purely a parametric surrogate for the  resonance integral. The $\psi_n$ are rational kernels and the $\omega_n(T)$ are temperature-dependent weights. In this formulation, Doppler broadening and other thermal effects are implicitly represented through the weights $\omega_n(T)$, while the nonlinear dependence on $\sigma_b$ is carried by the rational kernels $\psi_n(\sigma_b)$.

In this work we model only a systematic bias in the effective background $\sigma_b$, not in the temperature $T$. The NJOY reference tables are generated on prescribed temperature grid points, and our fitting reuses the same grid; thus $T$ acts as a fixed design variable rather than a noisy input, with thermal effects absorbed into the fitted weights $\omega_n(T)$. In contrast, $\sigma_b$ aggregates modeling choices (e.g., IR approximations and the use of $\lambda$) and composition-dependent inputs, so small model-form offsets naturally appear as a deterministic bias in $\sigma_b$. Accordingly, we now analyze how this bias is amplified by the nondecreasing-concave  response in $\sigma_b$ and by endpoint ill-conditioning, while treating $T$ as controlled and noise-free within the scope of this study.

Let $i$ index all sampled pairs $(\sigma_{b,i},\,T_i)$ on the design grid; the $i$-th Jacobian row is evaluated at that pair. 
With $f(\sigma_b,T)=\sum_{n}\omega_n(T)\,\psi_n(\sigma_b)$ and $\psi_n(\sigma_b)=\dfrac{\sigma_{a,n}\sigma_b}{\sigma_{a,n}+\sigma_b}$, the Jacobian entries are
\begin{equation}
\frac{\partial f_i}{\partial \omega_n}
= \psi_n(\sigma_{b,i}), \qquad
\frac{\partial f_i}{\partial \sigma_{a,n}}
= \omega_n(T_i)\,\frac{\sigma_{b,i}^2}{(\sigma_{a,n}+\sigma_{b,i})^2}.
\label{eq:df_partial}
\end{equation}

These formulas make clear how the two limiting regimes of $\sigma_b$ control both local sensitivity and the geometry of the Jacobian columns.

\noindent(i) Strong self-shielding ($\sigma_b\!\ll\!\sigma_{a,n}$).
Here
\begin{equation}
\frac{\partial f}{\partial \sigma_b}
= \sum_n \omega_n(T)\,
\frac{\sigma_{a,n}^2}{(\sigma_{a,n}+\sigma_b)^2}
\end{equation}
remains $O(1)$ and can be large, so any fixed abscissa offset $\Delta\sigma_b$ is relatively sensitivity-amplified:
\begin{equation}
e(\sigma_b,T)\;\approx\;
\left(\frac{\partial f}{\partial \sigma_b}\right)\,\Delta\sigma_b
\end{equation}
This produces large residual magnitudes at small $\sigma_b$ (high leverage), even when the model form is otherwise adequate.

\noindent(ii) Dilution  ($\sigma_b\!\gg\!\sigma_{a,n}$).
Now $\psi_n(\sigma_b)\!\approx\!\sigma_{a,n}$ and $\partial f/\partial\sigma_b=O(\sigma_b^{-2})$, so misfit magnitudes attenuate. 
However, in this limit the Jacobian columns become nearly constant across samples:
\begin{equation}
\frac{\partial f_i}{\partial \omega_n}
\;\approx\;
\sigma_{a,n}
\end{equation}
Consequently the weight-columns are almost parallel, so $J_\omega$ is nearly rank-one; when concatenated with the level block $J_a$, the full Jacobian $[\,J_\omega\ \ J_a\,]$ is driven toward rank deficiency. 
In the same limit,
\begin{equation}
\frac{\partial f_i}{\partial \sigma_{a,n}}
\;\approx\;
\omega_n(T_i)
\label{eq:df_dsigmaa_approx}
\end{equation}
Within each temperature block, this derivative remains essentially constant in the dilution limit. 
Under the assumption of weak Doppler broadening (as adopted here and consistent with Section~\ref{section:U238}), 
the temperature profiles of different blocks become further aligned. 
As a result, the column space of the sensitivity matrix is effectively compressed, leading to increased multicollinearity. 
Consequently, the column space collapses onto a few nearly identical directions, 
inflating the condition number of $J$ and rendering certain linear combinations of $\{\sigma_{a,n}\}$ poorly identifiable.

The analysis above is based on the response derived from the resonance integral, yet it still stands in the case of effective multigroup cross sections. Let us reconsider a response derived from the effective absorption cross section:
\begin{equation}
	f_{\mathrm{XG}}(\sigma_b; T) = \frac{\sum_{n\in g} \sigma_{a,n}\,\omega_n\,\dfrac{\sigma_b}{\sigma_{a,n}+\sigma_b}}
	{\sum_{n\in g} \omega_n\,\dfrac{\sigma_b}{\sigma_{a,n}+\sigma_b}}
	\label{eq:f_def}
\end{equation}
We again assume that the subgroup weights $\{\omega_n\}$ are non-negative and normalized, and that the subgroup levels $\{\sigma_{a,n}\}$ and the background cross section $\sigma_b$ are strictly positive.
It can be expressed as a weighted average $f(\sigma_b,T) = \sum_n p_n a_n$, with weights $p_n$ and terms $a_n$ defined as:
\begin{equation}
	a_n = \sigma_{a,n}, \quad
	p_n = \frac{\omega_n\,\dfrac{\sigma_b}{a_n+\sigma_b}}
	{\sum_k \omega_k\,\dfrac{\sigma_b}{a_k+\sigma_b}}
\end{equation}
The first derivative of $f$ with respect to $\sigma_b$ can then be written as:
\begin{equation}
	\begin{split}
		f' &= \sum_n p_n a_n g_n - \left(\sum_n p_n a_n\right) \left(\sum_k p_k g_k\right) \\
		&= \operatorname{Cov}_p(a_n, g_n) \\
		&= \tfrac{1}{2} \sum_{n,m} p_n p_m (a_n - a_m)(g_n - g_m) \\
		&\geq 0
	\end{split}
	\label{eq:f_prime_combined}
\end{equation}
where $g_n = \frac{1}{\sigma_b} - \frac{1}{a_n + \sigma_b}$. The equality to the third line utilizes the pairwise form of covariance, and the final inequality holds since $g_n$ is a monotonically increasing function of $a_n$ on its domain. Next, a direct differentiation yields
\begin{equation}
	f'' = \frac{2S}{W^3}\,\bigl(B^2 - C W \bigr).
\end{equation}
with
\begin{equation}\label{eq:defs}
\begin{aligned}
x_n &= \frac{1}{a_n+\sigma_b},  &\qquad
W &= \sum_k \omega_k x_k,      &\qquad
B &= \sum_k \omega_k x_k^{2}, \\
C   &= \sum_k \omega_k x_k^{3}, &\qquad
S   &= \sum_k \omega_k .
\end{aligned}
\end{equation}
Introduce
\begin{equation}
t_n = \sqrt{\omega_n} x_n^{3/2}, \quad v_n = \sqrt{\omega_n} x_n^{1/2}
\end{equation}
By the Cauchy-Schwarz inequality,
\begin{equation}
	\begin{split}
		B^2 &= \left( \sum_n t_n v_n \right)^2
		\leq \left( \sum_n t_n^2 \right)\left( \sum_n v_n^2 \right) = CW \\
		&\Rightarrow \quad f'' \le 0
	\end{split}
	\label{ref5}
\end{equation}
Equality holds if all $a_n$ are equal (i.e., $f$ is constant); otherwise, $f'' < 0$. Thus the response is nondecreasing and concave in $\sigma_b$, implying sensitivity concentrates at small $\sigma_b$ and diminishes in dilution---mirroring the surrogate before. 
While the underlying mechanisms differ slightly, they lead to the same practical challenges in the dilution limit. For the resonance-integral response, collinearity arises as the kernels $\psi_n(\sigma_b)$ themselves approach constants. For the effective-cross-section response, it is the entire function $f(\sigma_b, T)$ that flattens to a constant, making the distinct contributions of the $\{\sigma_{an}\}$ parameters inseparable.
Consequently, both responses exhibit similar practical difficulties: 
large leverage and heavy-tailed residuals in the strong self-shielding regime, 
and loss of parameter identifiability due to collinearity in the dilution limit. 
In addition, numerical artifacts near sharp resonances, temperature broadening of pointwise data, 
or incomplete convergence in fixed-source iterations can produce occasional large deviations 
at specific $(\sigma_b, T)$ points. 
Such outliers can dominate a quadratic loss and distort the estimation of subgroup parameters. 
These challenges---large leverage at small $\sigma_b$, collinearity in the dilution limit, 
and residual distortions caused by modeling bias---motivate the use of robust estimation methods.

\section{Robust estimation}

Robust statistics is concerned with developing methods that are resistant to deviations from the strict assumptions underlying classical parametric models. While classical procedures are optimal under ideal conditions like normality, their performance can be catastrophically poor in the presence of even slight violations. Real-world data frequently contain such deviations, which include the presence of gross errors (outliers), and the fact that the true data-generating distribution is often only an approximation of a theoretical model, for instance, exhibiting heavier tails than a perfect normal distribution.

To formalize and address these challenges, modern robust statistics was founded on two main theoretical pillars. The first, Huber's minimax theory \cite{huber1964robust}, formalizes departures from an idealized distribution $F_0$ via $\epsilon$-contamination neighborhoods
\begin{equation}
	\mathcal{P}_\epsilon(F_0) \;=\; \bigl\{\,F \;=\; (1-\epsilon)F_0 + \epsilon H \;:\; H \text{ arbitrary cdf}\,\bigr\}
\end{equation}
and seeks estimators that minimize the worst-case risk (e.g., asymptotic variance) over this neighborhood. The second, Hampel's infinitesimal approach \cite{hampel1968contributions}, assesses local robustness via the influence function (IF)
\begin{equation}
	\mathrm{IF}(x;T,F) \;=\; \lim_{t\downarrow 0}\,\frac{T\bigl((1-t)F+t\delta_x\bigr)-T(F)}{t}
\end{equation}
which quantifies the effect of a single outlier on the estimate. A key desideratum in this framework is a bounded influence function, ensuring that no single observation can have an arbitrarily large effect. Together, these theories provide a powerful foundation for robust estimation, with M-estimators emerging as a particularly flexible and effective class of solutions.

Building on this foundation, our work adopts an M-estimator with scale pre-estimation. This approach is chosen for its ability to effectively handle outliers while maintaining high efficiency. The residual for each data item $i$ is defined as
\begin{equation}\label{eq:residual-abs}
	e_i(\theta) \;=\; y_i - f_i(\theta),
\end{equation}
where $y_i$ is the observed datum. To ensure the method is scale---invariant---a crucial property for any robust procedure---a robust estimate of the residual scale, $\hat{s}$, is computed first using the median absolute deviation:
\begin{equation}
	\hat{s} = \frac{\text{median}_i \left| e_{i} - \text{median}(e_{i}) \right|}{0.6745}
\end{equation}
The denominator ensures that $\hat{s}$ is a consistent estimator for the standard deviation under a Gaussian model. The parameter vector is then estimated by solving the optimization problem:
\begin{equation}
	\hat{\theta} = \arg \min_{\theta} \sum_{i} \rho \left( \frac{e_{i}(\theta)}{\hat{s}} \right)
\end{equation}
For the $\rho$-function, we employ Tukey's bisquare function. This choice is motivated by its desirable "redescending" property, which completely rejects gross outliers rather than merely down-weighting them:
\begin{equation}
	\rho(u) = 
	\begin{cases} 
		\frac{c^2}{6} \left[ 1 - \left(1 - \left(\frac{u}{c}\right)^2\right)^3 \right] & \text{if } |u| \leq c \\
		\frac{c^2}{6} & \text{if } |u| > c
	\end{cases}
\end{equation}
The tuning constant is set to $c = 4.685$. This configuration achieves 95\% asymptotic efficiency for normally distributed data while completely discarding the influence of any observation whose standardized residual $|u|$ exceeds $c$.

While this formulation provides desirable statistical properties, it presents a significant computational challenge. The redescending nature of the Tukey's bisquare function makes the objective non-convex, creating a complex optimization landscape with multiple local minima. Furthermore, as detailed in Section \ref{Subgroup method}, the subgroup parameters are subject to physical and mathematical constraints that must be satisfied. To effectively tackle this constrained, non-linear optimization problem, a global search strategy is required. Therefore, as introduced in Section \ref{Differential evolution algorithm}, we employ the DE algorithm, a powerful meta-heuristic solver well-suited for such tasks.

\section{Differential evolution algorithm} \label{Differential evolution algorithm}
DE algorithm operates on a set of candidate solutions, referred to as a population. The algorithm iteratively improves these solutions through a combination of mutation, crossover, and selection operations. The key idea behind DE is to use the differences between randomly selected individuals in the population to guide the search process, thereby promoting exploration and exploitation of the search space. The basic operations of DE are as follows:
\paragraph{Initialization}
A population of $NP$ individuals $P = \{\vec{x}_1, \vec{x}_2, \dots, \vec{x}_{NP}\}$ is randomly generated within the lower $(\vec{x}_{\min})$ and upper $(\vec{x}_{\max})$ bounds of the search space. The $j$-th component of the $i$-th individual at generation $t=0$ is initialized as:
\begin{equation}
x_{i,j}^{(0)} = x_{\min,j} + \text{rand}(0,1) \cdot (x_{\max,j} - x_{\min,j})
\end{equation}
where $\text{rand}(0,1)$ is a uniform random number in $[0, 1]$.

\paragraph{Mutation}
For each individual $\vec{x}_i^{(t)}$ (the target vector) in the current population, a mutant vector $\vec{v}_i^{(t)}$ is generated. The most common strategy, "DE/rand/1", is defined as:
\begin{equation}
\vec{v}_i^{(t)} = \vec{x}_{r_1}^{(t)} + F \cdot (\vec{x}_{r_2}^{(t)} - \vec{x}_{r_3}^{(t)})
\end{equation}
where $r_1, r_2, r_3$ are mutually distinct integer indices randomly chosen from $\{1, \dots, NP\}$, and are different from the index $i$. The scaling factor $F$ controls the amplification of the differential variation.

\paragraph{Crossover}
To increase the diversity of the population, a trial vector $\vec{u}_i^{(t)}$ is formed by mixing the components of the mutant vector $\vec{v}_i^{(t)}$ with the target vector $\vec{x}_i^{(t)}$. Using exponential crossover, a starting point $n$ is randomly selected in $\{1, \dots, D\}$, and a length $L$ is determined based on the crossover probability $CR \in [0, 1]$. The trial vector is then formed as:
\begin{equation}
u_{i,j}^{(t)} = 
\begin{cases} 
	v_{i,j}^{(t)} & \text{for } j = \langle n \rangle_D, \langle n+1 \rangle_D, \dots, \langle n+L-1 \rangle_D \\
	x_{i,j}^{(t)} & \text{otherwise}
\end{cases}
\end{equation}
The length $L$ is chosen by repeatedly comparing a random number with $CR$, ensuring a block of consecutive components is inherited from the mutant vector. This method is particularly effective when variables have dependencies based on their ordering.

\paragraph{Selection}
A greedy selection scheme is applied. The trial vector $\vec{u}_i^{(t)}$ is compared to the target vector $\vec{x}_i^{(t)}$, and the one with the better (or equal) fitness value survives to the next generation $(t+1)$. For a minimization problem, this is expressed as:
\begin{equation}
\vec{x}_i^{(t+1)} = 
\begin{cases} 
	\vec{u}_i^{(t)} & \text{if } f(\vec{u}_i^{(t)}) \le f(\vec{x}_i^{(t)}) \\
	\vec{x}_i^{(t)} & \text{otherwise}
\end{cases}
\end{equation}
where $f(\cdot)$ is the objective function to be minimized. This ensures that the population's fitness never deteriorates.

Over the years, numerous DE variants and hybrid algorithms have been proposed to further enhance its performance and extend its applicability to different problem domains. In this paper, several such variants are employed. Since the focus of the study is not on algorithmic exploration, for comprehensive reviews of DE variants, readers may refer to \cite{ahmad2022differential,BILAL2020103479}.

\subsection{Constraint handling technique}
The determination of subgroup parameters requires adherence to physical and mathematical constraints (Section \ref{Modeling framework and limitations}). Ensuring feasibility is crucial, as numerical instabilities inherent in fitting procedures can yield unphysical results, such as negative parameters, a challenge noted in practice \cite{li2023analysis}. Our adoption of the Differential Evolution (DE) algorithm allows these constraints to be explicitly incorporated into the optimization process. To effectively manage these constraints within the DE framework, we employ the $\epsilon$-constrained method \cite{takahama2005constrained}. This technique modifies the DE selection process using a lexicographical comparison rule, prioritizing feasible solutions while navigating the search space towards optimality.

The general form of a constrained optimization problem is given as:
\begin{equation}
	\begin{aligned}
		& \text{Minimize:} && f(\mathbf{x}) \\
		& \text{subject to:} && g_j(\mathbf{x}) \leq 0, \quad j = 1, \dots, q \\
		&&& h_k(\mathbf{x}) = 0, \quad k = q+1, \dots, m
	\end{aligned}
\end{equation}
where $f(\mathbf{x})$ is the objective function, $g_j(\mathbf{x})$ are inequality constraints, and $h_k(\mathbf{x})$ are equality constraints for a solution vector $\mathbf{x} \in \mathbb{R}^n$.

The method first quantifies the degree of infeasibility for any solution using a constraint violation function, $\phi(\mathbf{x})$. This is typically defined as the sum of all individual constraint violations:
\begin{equation}
	\phi(\mathbf{x}) = \sum_{j=1}^{q} \max\{0, g_j(\mathbf{x})\} + \sum_{k=q+1}^{m} |h_k(\mathbf{x})|
\end{equation}
A solution $\mathbf{x}$ is considered feasible if and only if $\phi(\mathbf{x}) = 0$. The core of the technique is the subsequent use of the $\epsilon$-level comparison, which ranks solutions based on the ordered pair $(f(\mathbf{x}), \phi(\mathbf{x}))$. For any two solutions $\mathbf{x}_1$ and $\mathbf{x}_2$, with corresponding pairs $(f_1, \phi_1)$ and $(f_2, \phi_2)$, the relation $\leq_{\epsilon}$ is defined as:
\begin{equation}
	(f_1, \phi_1) \leq_{\epsilon} (f_2, \phi_2) \;\Leftrightarrow\;
	\begin{cases}
		f_1 \leq f_2, & \text{if } \phi_1, \phi_2 \leq \epsilon \\
		f_1 \leq f_2, & \text{if } \phi_1 = \phi_2 \\
		\phi_1 < \phi_2, & \text{otherwise}
	\end{cases}
\end{equation}
This rule establishes a clear "feasibility first" priority: if both solutions are acceptably close to the feasible region (i.e., their violations are below a threshold $\epsilon$) or have identical violation levels, the one with the better objective value is preferred. In all other cases, the solution with the smaller constraint violation is chosen, regardless of its objective value. A control mechanism often reduces $\epsilon$ over generations, gradually tightening the feasibility requirement.

To enhance search efficiency, especially for problems with equality constraints, a gradient-based mutation can be integrated \cite{Takahama2006}. It acts as a separate repair operation applied to an infeasible trial vector to guide it towards the feasible region using local gradient information.
The repair is based on a linear approximation of the constraint functions. First, a vector of all constraint functions, $C(\mathbf{x})$, and a corresponding vector of their violations, $\Delta C(\mathbf{x})$, are defined:
\begin{equation}
	C(\mathbf{x}) = [g_1(\mathbf{x}), \dots, g_q(\mathbf{x}), h_{q+1}(\mathbf{x}), \dots, h_m(\mathbf{x})]^T
\end{equation}
\begin{equation}
	\Delta C(\mathbf{x}) = [\max\{0, g_1(\mathbf{x})\}, \dots, \max\{0, g_q(\mathbf{x})\}, h_{q+1}(\mathbf{x}), \dots, h_m(\mathbf{x})]^T
\end{equation}
The goal is to find a corrective step, $\Delta\mathbf{x}$, that nullifies the current constraint violation. This is achieved by solving the linear system $\nabla C(\mathbf{x}) \Delta\mathbf{x} = -\Delta C(\mathbf{x})$, where $\nabla C(\mathbf{x})$ is the $m \times n$ Jacobian matrix of the constraint functions:
\begin{equation}
	\nabla C(\mathbf{x}) =
	\begin{pmatrix}
		\frac{\partial g_1}{\partial x_1} & \cdots & \frac{\partial g_1}{\partial x_n} \\
		\vdots & \ddots & \vdots \\
		\frac{\partial g_q}{\partial x_1} & \cdots & \frac{\partial g_q}{\partial x_n} \\
		\frac{\partial h_{q+1}}{\partial x_1} & \cdots & \frac{\partial h_{q+1}}{\partial x_n} \\
		\vdots & \ddots & \vdots \\
		\frac{\partial h_m}{\partial x_1} & \cdots & \frac{\partial h_m}{\partial x_n}
	\end{pmatrix}
\end{equation}
Since the Jacobian is generally not square or invertible, the Moore-Penrose pseudoinverse, denoted $(\cdot)^+$, is used to calculate the correction:
\begin{equation}
	\Delta\mathbf{x} = -[\nabla C(\mathbf{x})]^{+} \Delta C(\mathbf{x})
\end{equation}
The infeasible trial vector, $\mathbf{x}_{\text{trial}}$, is then repaired by:
\begin{equation}
	\mathbf{x}_{\text{repaired}} = \mathbf{x}_{\text{trial}} + \Delta\mathbf{x}
\end{equation}
This operation is typically applied conditionally with a probability $P_g$ to a trial vector if it is not $\epsilon$-feasible. This strategy uses local gradient information to efficiently improve solution feasibility without disrupting the global search behavior of the DE algorithm.

\section{Results} \label{Results}

\begin{table*}[t]
    \centering
    \begin{tabular}{lllllll}
        \toprule
        Case & Identification & EALF(eV) & Q & $k_{exp}$  & Sensitive Nuclides\\
        \midrule

        1 & TRX1 &         & 0.0946 & 1.0000  & $^{1}\mathrm{H},\; ^{235}\mathrm{U},\; ^{238}\mathrm{U},\; ^{16}\mathrm{O}$\\
        2 & TRX2 &         & 0.0693 & 1.0000  & $^{1}\mathrm{H},\; ^{235}\mathrm{U},\; ^{238}\mathrm{U},\; ^{16}\mathrm{O}$\\

        3 & BAPL1 & \multirow{3}{*}{~} & 0.078 & 1.0000  & $^{1}\mathrm{H},\; ^{235}\mathrm{U},\; ^{238}\mathrm{U},\; ^{16}\mathrm{O}$\\
        4 & BAPL2 &         & 0.07   & 1.0000  & $^{1}\mathrm{H},\; ^{235}\mathrm{U},\; ^{238}\mathrm{U},\; ^{16}\mathrm{O}$\\
        5 & BAPL3 &         & 0.057  & 1.0000  & $^{1}\mathrm{H},\; ^{235}\mathrm{U},\; ^{238}\mathrm{U},\; ^{16}\mathrm{O}$\\

        6 & BNLuma1 & \multirow{6}{*}{~} & 0.646 & 1.0000  & $^{1}\mathrm{H},\; ^{235}\mathrm{U},\; ^{238}\mathrm{U},\; ^{16}\mathrm{O}$\\
        7 & BNLuma2 &         & 0.712 & 1.0000  & $^{1}\mathrm{H},\; ^{235}\mathrm{U},\; ^{238}\mathrm{U},\; ^{16}\mathrm{O}$\\
        8 & BNLuma4 &         & 0.602 & 1.0000  & $^{1}\mathrm{H},\; ^{235}\mathrm{U},\; ^{238}\mathrm{U},\; ^{16}\mathrm{O}$\\
        9 & BNLuma5 &         & 0.717 & 1.0000  & $^{1}\mathrm{H},\; ^{235}\mathrm{U},\; ^{238}\mathrm{U},\; ^{16}\mathrm{O}$\\
        10 & BNLuma6 &        & 0.542 & 1.0000  & $^{1}\mathrm{H},\; ^{235}\mathrm{U},\; ^{238}\mathrm{U},\; ^{16}\mathrm{O}$\\
        11 & BNLuma7 &        & 0.797 & 1.0000  & $^{1}\mathrm{H},\; ^{235}\mathrm{U},\; ^{238}\mathrm{U},\; ^{16}\mathrm{O}$\\

        12 & HST13.1 & 0.0327 & \multirow{4}{*}{~} & 1.2192  & $^{1}\mathrm{H},\; ^{235}\mathrm{U},\; ^{16}\mathrm{O}$\\
        13 & HST13.2 & 0.0341 &         & 1.2143  & $^{1}\mathrm{H},\; ^{235}\mathrm{U},\; ^{10}\mathrm{B},\; ^{16}\mathrm{O}$\\
        14 & HST13.3 & 0.0355 &         & 1.2110  & $^{1}\mathrm{H},\; ^{235}\mathrm{U},\; ^{10}\mathrm{B},\; ^{16}\mathrm{O}$\\
        15 & HST13.4 & 0.0362 &         & 1.2081  & $^{1}\mathrm{H},\; ^{235}\mathrm{U},\; ^{10}\mathrm{B},\; ^{16}\mathrm{O}$\\

        16 & PST11.10 & 0.055  & \multirow{5}{*}{~} & 1.4692  & $^{239}\mathrm{Pu},\; ^{1}\mathrm{H},\; ^{113}\mathrm{Cd},\; ^{16}\mathrm{O},\; ^{14}\mathrm{N},\; ^{240}\mathrm{Pu}$\\
        17 & PST11.11 & 0.0584 &         & 1.4911  & $^{239}\mathrm{Pu},\; ^{1}\mathrm{H},\; ^{14}\mathrm{N},\; ^{113}\mathrm{Cd},\; ^{16}\mathrm{O},\; ^{240}\mathrm{Pu}$\\
        18 & PST11.12 & 0.0536 &         & 1.4759  & $^{239}\mathrm{Pu},\; ^{1}\mathrm{H},\; ^{113}\mathrm{Cd},\; ^{16}\mathrm{O},\; ^{14}\mathrm{N},\; ^{240}\mathrm{Pu}$\\
        19 & PST21.7  & 0.061  &         & 1.6021  & $^{1}\mathrm{H},\; ^{239}\mathrm{Pu},\; ^{16}\mathrm{O},\; ^{240}\mathrm{Pu}$\\
        20 & PST21.8  & 0.324  &         & 1.6468  & $^{1}\mathrm{H},\; ^{239}\mathrm{Pu},\; ^{16}\mathrm{O},\; ^{240}\mathrm{Pu}$\\

        \bottomrule
    \end{tabular}
	\caption{Benchmarks}
    \label{benchmarks}
\end{table*}

To evaluate the performance of the proposed method, a set of well-documented benchmarks is selected, as listed in Table~\ref{benchmarks}. Cases~1--2 are light-water moderated, low-enriched uranium metal hexagonal lattices
($^{235}$U~$\approx$~1.3~wt\%), which differ in lattice pitch (1.806 cm and 2.174 cm). Cases~3--5 are light-water moderated, low-enriched uranium dioxide lattices in a uniform hexagonal array. Across these subcases, the series spans three hexagonal pitches of 1.5578~cm, 1.6523~cm, and 1.8057~cm. Cases~6--11 are light-water moderated, low-enriched uranium metal lattices in uniform hexagonal arrays. Four enrichments were considered (1.016, 1.027, 1.143, and 1.299~wt\% $^{235}$U), with variations in lattice pitch across the BNL series. Cases~12--15 are bare spheres of enriched uranyl nitrate solution based on 93.2~wt\% $^{235}$U, contained in an aluminum spherical vessel with a diameter of ~69.2~cm. The H/$^{235}$U ranges from $\sim$972 to $\sim$1378. The last three out of four cases include borated solution. Cases~16--18 are bare plutonium nitrate solution spheres with a $^{240}$Pu content of 4.17--4.20~wt\%, housed in stainless-steel spherical vessels of radii 40.6~cm and 45.7~cm. Cases~19--20 are bare plutonium nitrate solution spheres with $^{240}$Pu at 4.57~wt\%, contained in a stainless-steel sphere of radius 19.5~cm.

Using a 394-group energy structure, the resonance self-shielding is treated for the key resonance nuclides $\mathrm{^{235}U}$, $\mathrm{^{238}U}$, $\mathrm{^{239}Pu}$, and $\mathrm{^{240}Pu}$, with the resulting parameters subsequently used by the \textsc{ALPHA} code \cite{li2021investigation,liang2020overlapping,song2019implementation,song2020implementation} to obtain the transport results.
Both the RE method and the OLS method are applied to the benchmark problems, and the numerical results are compared against experimental data \cite{alter1974cross,NEA:ICSBEP:2006}.
The relative errors are summarized in Table~\ref{relative_erro}.

\subsection{Impact of $\mathrm{^{238}U}$ Physics and Multicollinearity}
\label{section:U238}

For cases~12--20, the transport results obtained with RE and OLS are identical. 
Cross-referencing the sensitivity analysis (Table~\ref{benchmarks}) with the benchmark results (Table~\ref{relative_erro}) reveals that the benchmarks showing noticeable improvement are exclusively those involving $\mathrm{^{238}U}$ as a sensitive nuclide.  This consistency strongly suggests that $\mathrm{^{238}U}$ plays a central role in the observed discrepancy, a role that can be attributed to a combination of its microscopic resonance structure and its macroscopic abundance in the reactor core. The first factor is the distinctive resonance structure of \(\mathrm{^{238}U}\), which stands in sharp contrast to that of fissile nuclides. As a fertile, even-even nucleus, \(\mathrm{^{238}U}\) exhibits resonances that are widely spaced, reflecting a smaller accessible level density at the relevant low excitation energies. Two mechanisms dominate: (i) nucleon pairing suppresses the intrinsic level density in this region, and (ii) spin-parity selection from a \(0^+\) target severely limits the number of allowed low-energy entrance sequences. Together these effects produce the sparse, well-separated resonances characteristic of \(\mathrm{^{238}U}\).
This distinct topography---composed of sparse, narrow resonances separated by extensive, flat valleys---gives rise to an extremely pronounced self-shielding effect \cite{cullen2019importance} and makes it more probable that a discrete energy group will sample regions of low, flat cross-section, which may be the deep valleys between major peaks or the saturated, far-field tails of resonances.
This structural feature creates the conditions for severe multicollinearity through two compounding physical effects: (i) rapid dilution-induced  saturation, where the self-shielding effect quickly vanishes and the response saturates at its infinitely dilute value. This occurs because in these low, flat cross-section regions, the nuclide's resonance cross-section ($\sigma_a$) becomes relatively insignificant, causing it to approximate the behavior of other non-resonant materials in the mixture (which contribute $\sigma_0$). While the specific mechanism varies---saturating individual kernel functions in one model versus flattening the entire response function in another, as detailed in Section~\ref{Modeling framework and limitations}---the outcome is the same; and (ii) weak Doppler broadening in these flat regions, which renders the response value equally insensitive to changes in temperature ($T$). Together, these effects dramatically reduce the variance of the reference data for such groups across the entire $(\sigma_b, T)$ fitting space.
Consequently, the distinct contributions of the subgroup parameters become inseparable, leading to the severe multicollinearity that destabilizes OLS solutions.

\begin{table}[htbp]
    \centering
    \begin{tabular}{lcc} 
        \toprule
        Case & \multicolumn{2}{c}{Relative error (pcm)}\\
        \cmidrule(l{2pt}r{2pt}){2-3}
        & OLS & RE \\
        \midrule

        1 & -520 & -440 \\
        2 & -240 & -190 \\
        3 & -370 & -290 \\
        4 & -270 & -210 \\
        5 & -180 & -130 \\
        6 & -160 & -140 \\
        7 & -670 & -650 \\
        8 & -30  & 50  \\
        9 & -700 & -620 \\
        10 & 710 & 770 \\
        11 & -430 & -400 \\
        12 & -541 & -541 \\
        13 & -576 & -576 \\
        14 & -933 & -933 \\
        15 & -745 & -745 \\
        16 & -299 & -299 \\
        17 & -664 & -664 \\
        18 & -630 & -630 \\
        19 & -87  & -87  \\
        20 & -279 & -279 \\

        \bottomrule
    \end{tabular}
	\caption{Relative error of $k_{calc}$ with respect to experimental data}
    \label{relative_erro}
\end{table}

While other fertile nuclides, notably $\mathrm{^{240}Pu}$, share similar microscopic resonance characteristics, the numerical impact of $\mathrm{^{238}U}$ is uniquely dominant due to both the pervasive nature of its resonance structure and its overwhelming abundance. The primary structural difference lies in the energy distribution of their resonances. The influence of $\mathrm{^{240}Pu}$ is heavily localized around its single, large resonance near 1~eV, whereas $\mathrm{^{238}U}$ possesses a series of strong resonances distributed throughout the entire epithermal range. This pervasive structure affects a much broader portion of the neutron slowing-down spectrum. The difference in microscopic spectral impact is then dramatically amplified by their disparate macroscopic concentrations. Constituting over 90\% of typical fuel, the number density of $\mathrm{^{238}U}$ is orders of magnitude greater than that of $\mathrm{^{240}Pu}$, a secondary transmutation product. Since self-shielding is governed by the macroscopic cross-section ($\Sigma = N\sigma$), this massive concentration elevates $\mathrm{^{238}U}$'s pervasive resonance structure into a dominant, first-order effect. Thus, it is the combination of its pervasive microscopic resonance structure and its dominant macroscopic abundance that makes $\mathrm{^{238}U}$ the principal source of the observed numerical instabilities.

\begin{figure}[htbp]
    \centering
    \includegraphics[width=\columnwidth]{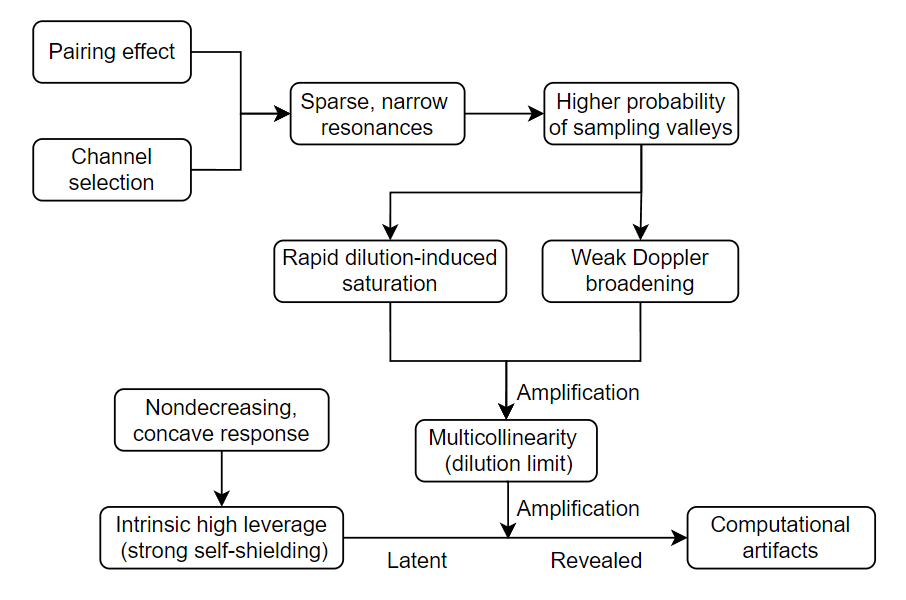}

    \caption{Causal chain of threshold-like artifacts}
    \label{fig:mechanism}
\end{figure}

\subsection{Systematic OLS tilt under an inferred $\sigma_b$ offset}

For benchmarks sensitive to $^{238}$U, OLS yields systematically lower reactivity than RE (Table~\ref{relative_erro}), indicating a larger absorption-type response $f(\sigma_b,T)$ under OLS. This follows from the structure of the model. As shown in Section~\ref{Modeling framework and limitations}, $f$ is nondecreasing and concave in $\sigma_b$ ($f'\!\ge0$, $f''\!\le0$). Using an effective background $\sigma_b+\Delta$ produces residuals
\begin{equation}
r_i(\theta)=f(\sigma_{b,i}+\Delta;\theta)-f(\sigma_{b,i})\approx f'(\sigma_{b,i})\,\Delta,
\end{equation}
which share the sign of $\Delta$ and attain their largest magnitudes at small $\sigma_b$ where $f'$ is maximal. Minimizing the squared error therefore concentrates influence in the strong self-shielding region and drives parameters to raise $f$ there. The observed upward shift of $f$ implies a negative effective offset in the high-leverage region, $\Delta<0$.

In addition, the two responses deviate by different amounts under this tilt. In the strong self-shielding limit ($\sigma_b\!\ll\!\sigma_{a,n}$), the XG response reduces to a weighted harmonic mean,
\begin{equation}
f_{\mathrm{XG}}(\sigma_b,T)\;\approx\;\Big(\sum_{n}\tfrac{\omega_n}{\sigma_{a,n}}\Big)^{-1},
\end{equation}
so shifting weight toward larger internal levels (or raising those levels) increases $f_{\mathrm{XG}}$ efficiently; under the same OLS tilt, XG therefore exhibits a larger upward deviation. By contrast, the RI response satisfies
\begin{equation}
f_{\mathrm{RI}}(\sigma_b,T)\;=\;\sum_n \omega_n\,\frac{\sigma_{a,n}\sigma_b}{\sigma_{a,n}+\sigma_b}
\;\approx\;\sigma_b,
\end{equation}
and is nearly insensitive to $(\omega_n,\sigma_{a,n})$ in this limit. Its upward adjustment occurs mainly over small---to---moderate $\sigma_b$ where $\partial f_{\mathrm{RI}}/\partial \sigma_{a,n}>0$, leading to a weaker net deviation. In the dilution limit, both responses flatten and offer little counter-penalty, so upward adjustments made at low $\sigma_b$ persist, with XG typically showing the larger bias. Overall, the disparity arises because the low-$\sigma_b$ samples both dominate the loss and, for XG, provide effective parameter directions to raise the response, whereas RI offers little such leverage in that regime.

\begin{figure}[htbp]
    \centering
    \includegraphics[width=\columnwidth]{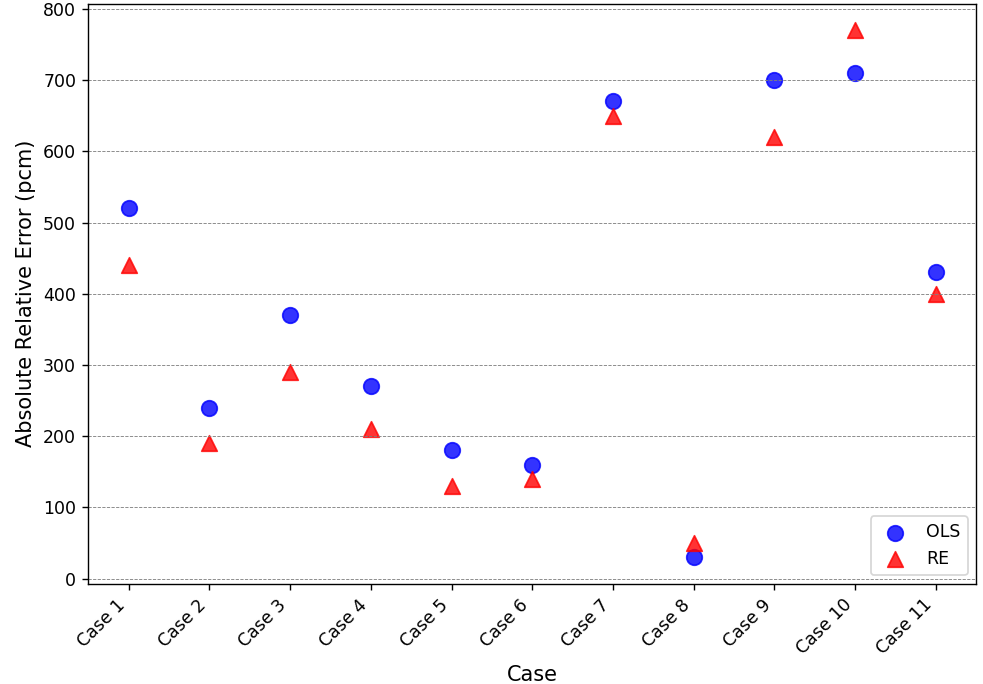}

    \caption{Magnitude of the relative error of $k_{calc}$ with respect to experimental data in cases sensitive to $^{238}$U}
    \label{fig:comparison}
\end{figure}

\subsection{Exceptions and Compensating Biases}

While RE improves most $^{238}$U---sensitive benchmarks, Cases~8 and~10 are exceptions where RE yields less accurate $k$ (Figure~\ref{fig:comparison}). Across the suite, OLS systematically predicts lower $k$ than RE, consistent with an absorption-increasing tilt from the self-shielding fit; this is an interpretation of a repeated pattern, not a proven mechanism. For Cases~8 and~10, a plausible cause is compensating bias across stages: if transport introduces a positive reactivity bias (e.g., underpredicted leakage), the OLS tilt (negative reactivity) can cancel part of it, yielding an apparently better $k$. When RE reduces the self-shielding tilt, the cancellation vanishes and the latent transport bias becomes visible, increasing the deviation. Similar cancellations have been noted for resonance treatments across groups or regions~\cite{rosier2025analysis}, and can likewise occur across computational stages.

\section{Conclusions}
We introduced a robust subgroup method that integrates principles from RE with a DE algorithm to address the inherent limitations of the OLS method for determining subgroup parameters.  Validation demonstrated that this approach successfully mitigated a systematic absorption bias, yielding more physically consistent and globally accurate subgroup parameters, particularly for benchmarks sensitive to the characteristic resonance structure of key fertile nuclide $^{238}$U. 

Our analysis offers a deeper perspective on the challenges in high-fidelity simulation. The systematic absorption bias is a computational artifact arising when a small modeling bias is amplified by the concave response function, creating high-leverage residuals in the strong self-shielding regime. Consequently, the quadratic loss function allows these few residuals to exert a disproportionate influence on the regression, thereby skewing the parameter estimates. However, this instability appears to be dramatically amplified into an observable, dominant error only when triggered by severe multicollinearity---an effect we trace back to the specific resonance structural properties of certain nuclides under fine-group discretization. Thus, the systematic absorption bias, which may otherwise be latent, manifests as a critical, dominant error when the simulation's fidelity crosses a threshold where the OLS method becomes unstable. 

Looking ahead, this study suggests several promising avenues for future work. (i)~A key next step is to validate our hypothesis of fidelity-dependent instability. A systematic study should confirm if the significant OLS-RE discrepancy vanishes in coarse-group structures, as predicted. Such a study would confirm fine-group-induced multicollinearity as the trigger mechanism and help define the fidelity threshold beyond which robust methods become essential.
(ii)~The instances where OLS results appeared more accurate due to a likely cancellation of errors highlight the need for a holistic approach, motivating efforts to extend this robust framework across the entire simulation chain. (iii)~Finally, a direction lies in enhancing the method's fidelity by employing a larger number of subgroups, i.e., expanding the set of rational kernels used to approximate the physical quantities. However, this would likely transition the parameter estimation into a high-dimensional setting, introducing significant challenges related to computational tractability and the fundamental stability of statistical estimators. The burgeoning field of high-dimensional robust statistics \cite{hastie2015statistical,loh2024theoretical} offers a principled framework for navigating the associated challenges, providing a roadmap for extending the proposed method.

In conclusion, this work establishes the robust subgroup method as a reliable foundation for resonance treatment by directly confronting a key computational artifact. Our analysis identifies this artifact as a systematic absorption bias, resulting from a numerical instability triggered by high-fidelity, fine-group discretization. This framework therefore offers a principled path to help ensure that increased fidelity translates to genuine accuracy, rather than new computational
artifacts.

\section*{CRediT authorship contribution statement}

\textbf{Beichen Zheng:} Conceptualization, Methodology, Software, Formal analysis, Writing - original draft, Writing - review \& editing. 
\textbf{Ying Chen:} Validation.
\textbf{Lili Wen:} Data curation.
\textbf{Xiaofei Wu:} Project administration.

\bibliographystyle{elsarticle-harv}
\bibliography{RSM-refs.bib}

@article{huber1964robust,
  title = {Robust Estimation of a Location Parameter},
  author = {Huber, Peter J.},
  year = {1964},
  journal = {Annals of Mathematical Statistics},
  volume = {35},
  number = {1},
  pages = {73--101},
}

@article{rosier2025analysis,
  title={An analysis of a coarse-group subgroup method based on the physical probability tables in apollo3{\textregistered}},
  author={Rosier, Emeline and Mao, Li and Sanchez, Richard and Leal, Luiz and Zmijarevic, Igor},
  journal={Nuclear Science and Engineering},
  volume={199},
  number={sup1},
  pages={S121--S134},
  year={2025},
  publisher={Taylor \& Francis}
}

@article{sugimura2007resonance,
  title = {Resonance Treatment Based on Ultra-fine-group Spectrum Calculation in the AEGIS Code},
  author = {Sugimura, Naoki and Yamamoto, Akio},
  year = {2007},
  journal = {Journal of Nuclear Science and Technology},
  volume = {44},
  number = {7},
  pages = {958--966},
}

@techreport{askew1966general,
  title={General description of the lattice code {WIMS}},
  author={Askew, J.R. and Fayers, F.J. and Kemshell, P.B.},
  year={1966},
  institution={Atomic Energy Establishment, Winfrith, Eng.}
}

@book{stamm1983methods,
  title={Methods of Steady-State Reactor Physics in Nuclear Design},
  author={Stamm'ler, Rudi J. J. and Abbate, M{\'a}ximo Julio},
  publisher={Academic Press},
  year={1983}
}

@techreport{ishiguro1971peaco,
  author = {Ishiguro, Y. and Takano, H.},
  title = {PEACO: A code for calculation of group constant of resonance energy region in heterogeneous systems},
  institution = {Japan Atomic Energy Research Institute},
  number  = {JAERI-1},
  year  = {1971}
}

@article{levitt1972probability,
  title={The probability table method for treating unresolved neutron resonances in Monte Carlo calculations},
  author={Levitt, Leo B},
  journal={Nuclear Science and Engineering},
  volume={49},
  number={4},
  pages={450--457},
  year={1972},
  publisher={Taylor \& Francis}
}

@article{cullen1974application,
  title={Application of the probability table method to multigroup calculations of neutron transport},
  author={Cullen, Dermott E},
  journal={Nuclear Science and Engineering},
  volume={55},
  number={4},
  pages={387--400},
  year={1974},
  publisher={Taylor \& Francis}
}

@article{nikolaev1976comments,
  title={Comments on the probability table method},
  author={Nikolaev, MN},
  journal={Nuclear Science and Engineering},
  volume={61},
  number={2},
  pages={286--287},
  year={1976},
  publisher={Taylor \& Francis}
}

@article{hebert2002computing,
  title={Computing moment-based probability tables for self-shielding calculations in lattice codes},
  author={H{\'e}bert, Alain and Coste, Mireille},
  journal={Nuclear Science and Engineering},
  volume={142},
  number={3},
  pages={245--257},
  year={2002},
  publisher={Taylor \& Francis}
}

@article{chiba2006improvement,
  title={Improvement of moment-based probability table for resonance self-shielding calculation},
  author={Chiba, Go and Unesaki, Hironobu},
  journal={Annals of Nuclear Energy},
  volume={33},
  number={13},
  pages={1141--1146},
  year={2006},
  publisher={Elsevier}
}

@article{li2021improved,
  title={Improved subgroup method coupled with particle swarm optimization algorithm for intra-pellet non-uniform temperature distribution problem},
  author={Li, Song and Zhang, Qian and Zhang, Zhijian and Zhao, Qiang and Liang, Liang},
  journal={Annals of Nuclear Energy},
  volume={153},
  pages={108070},
  year={2021},
  publisher={Elsevier}
}

@article{safarzadeh2015resonance,
  title={Resonance self-shielding calculation using subgroup method and ABC algorithm},
  author={Safarzadeh, Omid and Shirani, AS and Minuchehr, A},
  journal={Progress in Nuclear Energy},
  volume={78},
  pages={303--309},
  year={2015},
  publisher={Elsevier}
}

@techreport{ribon1986probability,
  title={Probability tables and gauss quadrature: application to neutron cross-sections in the unresolved energy range},
  author={Ribon, P and Maillard, JM and others},
  year={1986},
  institution={CEA Centre d'Etudes Nucleaires de Saclay, Inst.}
}

@book{hampel1968contributions,
  title={Contributions to the theory of robust estimation},
  author={Hampel, Frank Rudolf},
  year={1968},
  publisher={University of California, Berkeley}
}

@article{osaba2021tutorial,
  title={A tutorial on the design, experimentation and application of metaheuristic algorithms to real-world optimization problems},
  author={Osaba, Eneko and Villar-Rodriguez, Esther and Del Ser, Javier and Nebro, Antonio J and Molina, Daniel and LaTorre, Antonio and Suganthan, Ponnuthurai N and Coello, Carlos A Coello and Herrera, Francisco},
  journal={Swarm and Evolutionary Computation},
  volume={64},
  pages={100888},
  year={2021},
  publisher={Elsevier}
}

@article{BILAL2020103479,
  title = {Differential Evolution: A review of more than two decades of research},
  journal = {Engineering Applications of Artificial Intelligence},
  volume = {90},
  pages = {103479},
  year = {2020},
  author = {Bilal and Pant, Millie and Zaheer, Hira and Garcia-Hernandez, Laura and Abraham, Ajith},
  publisher={Elsevier}
}

@article{ahmad2022differential,
  title={Differential evolution: A recent review based on state-of-the-art works},
  author={Ahmad, Mohamad Faiz and Isa, Nor Ashidi Mat and Lim, Wei Hong and Ang, Koon Meng},
  journal={Alexandria Engineering Journal},
  volume={61},
  number={5},
  pages={3831--3872},
  year={2022}, 
  publisher={Elsevier}
}

@techreport{storn1995differential,
  title={Differrential evolution-a simple and efficient adaptive scheme for global optimization over continuous spaces},
  author={Storn, Rainer},
  institution={International Computer Science Institute},
  year={1995},
  type={Technical report},
  volume={11}
}

@techreport{Stammler1998,
  author = {Stamm'ler, R. J.},
  title = {HELIOS methods},
  institution = {Studsvik Scandpower},
  year = {1998}
}

@techreport{kim2016subgr,
  title={SUBGR: A Program to Generate Subgroup Data for the Subgroup Resonance Self-Shielding Calculation},
  author={Kim, Kang Seog},
  year={2016},
  institution={Oak Ridge National Lab.(ORNL), Oak Ridge, TN (United States)}
}

@article{joo2009subgroup,
  title={Subgroup weight generation based on shielded pin-cell cross section conservation},
  author={Joo, Han Gyu and Kim, Gwan Young and Pogosbekyan, Leonid},
  journal={Annals of Nuclear Energy},
  volume={36},
  number={7},
  pages={859--868},
  year={2009},
  publisher={Elsevier}
}

@inproceedings{takahama2005constrained,
  title={Constrained optimization by {$\varepsilon$} constrained particle swarm optimizer with {$\varepsilon$}-level control},
  author={Takahama, Tetsuyuki and Sakai, Setsuko},
  booktitle={Soft Computing as Transdisciplinary Science and Technology: Proceedings of the fourth IEEE International Workshop WSTST'05},
  pages={1019--1029},
  year={2005},
  organization={Springer}
}

@inproceedings{Takahama2006,
  author = {Takahama, Tetsuyuki and Sakai, Setsuko},
  title  = {Constrained Optimization by the {$\varepsilon$} Constrained Differential Evolution with Gradient-Based Mutation and Feasible Elites},
  booktitle = {2006 IEEE International Conference on Evolutionary Computation (CEC 2006)},
  year = {2006},
  pages  = {308--315},
  publisher = {IEEE}
}

@techreport{cullen2019importance,
  title        = {The importance of resonance self-shielding},
  author       = {Cullen, Dermott E.},
  year         = {2019},
  institution  = {International Nuclear Data Committee, IAEA},
  number       = {INDC(NDS)-0778},
}

@inproceedings{casal1991helios,
  title={HELIOS: geometric capabilities of a new fuel-assembly program},
  author={Casal, J},
  booktitle={Proc. Int. Topical Meeting on Advances in Mathematics, Computations and Reactor Physics, Pittsburgh, Pa., USA, April 28-May 2, 1991},
  volume={2},
  pages={10--2},
  year={1991}
}

@techreport{alter1974cross,
  title={Cross section evaluation working group benchmark specifications},
  author={Alter, H and Kidman, RB and LaBauve, R and Prostik, R and Zolotar, BA},
  year={1974},
  institution={Brookhaven National Laboratory (BNL), Upton, NY (United States)}
}

@techreport{NEA:ICSBEP:2006,
  author = {{Nuclear Energy Agency}},
  title  = {International Handbook of Evaluated Criticality Safety Benchmark Experiments},
  institution = {Organisation for Economic Co-operation and Development (OECD/NEA)},
  year = {2006},
  type = {Technical Report},
  number = {NEA/NSC/DOC(95)04/I}
}

@article{li2021investigation,
  title={Investigation of the efficiency optimization for the improved subgroup resonance self-shielding treatment on the GPU platform},
  author={Li, Song and Zhang, Qian and Zhang, Zhijian and Song, Peitao and Liang, Liang and Zhao, Qiang},
  journal={Annals of Nuclear Energy},
  volume={159},
  pages={108318},
  year={2021},
  publisher={Elsevier}
}

@article{liang2020overlapping,
  title={Overlapping communication and computation of GPU/CPU heterogeneous parallel spatial domain decomposition MOC method},
  author={Liang, Liang and Zhang, Qian and Song, Peitao and Zhang, Zhijian and Zhao, Qiang and Wu, Hongchun and Cao, Liangzhi},
  journal={Annals of Nuclear Energy},
  volume={135},
  pages={106988},
  year={2020},
  publisher={Elsevier}
}

@article{song2019implementation,
  title={Implementation and performance analysis of the massively parallel method of characteristics based on GPU},
  author={Song, Peitao and Zhang, Zhijian and Liang, Liang and Zhang, Qian and Zhao, Qiang},
  journal={Annals of Nuclear Energy},
  volume={131},
  pages={257--272},
  year={2019},
  publisher={Elsevier}
}

@article{song2020implementation,
  title={Implementation of the CPU/GPU hybrid parallel method of characteristics neutron transport calculation using the heterogeneous cluster with dynamic workload assignment},
  author={Song, Peitao and Zhang, Zhijian and Zhang, Qian and Liang, Liang and Zhao, Qiang},
  journal={Annals of Nuclear Energy},
  volume={135},
  pages={106957},
  year={2020},
  publisher={Elsevier}
}

@article{hastie2015statistical,
  title={Statistical learning with sparsity},
  author={Hastie, Trevor and Tibshirani, Robert and Wainwright, Martin},
  journal={Monographs on statistics and applied probability},
  volume={143},
  number={143},
  pages={8},
  year={2015}
}

@article{li2023analysis,
  title={Analysis of the fine-mesh subgroup method and its feasible improvement},
  author={Li, Song and Zhang, Qian and Liu, Lei and Zhang, Yongfa and Hao, Jianli and Wang, Xiaolong and Jiang, Lizhi and Liu, Xiaoya},
  journal={Frontiers in Energy Research},
  volume={10},
  pages={1036063},
  year={2023},
  publisher={Frontiers Media SA}
}

@article{yin2025study,
  title={Study on the subgroup method for the resonance self-shielding calculations in VITAS},
  author={Yin, Han and Zhang, Qian and Liu, Xiaojing and He, Hui and Zhang, Tengfei},
  journal={Nuclear Engineering and Design},
  volume={432},
  pages={113801},
  year={2025},
  publisher={Elsevier}
}

@article{zhao2025research,
  title={Research of the few-group cross-sections calculation based on the software package MOSASAUR},
  author={Zhao, Chen and Wang, Lianjie and Lou, Lei and Zhang, Bin and Zhang, Longlong and Xia, Bangyang},
  journal={Progress in Nuclear Energy},
  volume={188},
  pages={105884},
  year={2025},
  publisher={Elsevier}
}

@article{loh2024theoretical,
  title={A theoretical review of modern robust statistics},
  author={Loh, Po-Ling},
  journal={Annual Review of Statistics and Its Application},
  volume={12},
  year={2024},
  publisher={Annual Reviews}
}

@article{goldstein1962theory,
  title={Theory of resonance absorption of neutrons},
  author={Goldstein, Rubin and Cohen, E Richard},
  journal={Nuclear Science and Engineering},
  volume={13},
  number={2},
  pages={132--140},
  year={1962},
  publisher={Taylor \& Francis}
}

\end{document}